# Opto-thermoelectric trapping of Fluorescent Nanodiamonds on Plasmonic Nanostructures

Ashutosh Shukla[1], Sunny Tiwari[1,2], Ayan Majumder[3], Kasturi Saha[3], and G.V.Pavan Kumar[1,*],

[1]*Department of Physics, Indian Institute of Science Education and Research, Pune, India*
[2]*(present address) Department of Physics, Clarendon Laboratory, University of Oxford, Oxford OX1 3PU, United Kingdom*
[3]*Department of Electrical Engineering, Indian Institute of Technology, Bombay, India*
*\*Corresponding author: pavan@iiserpune.ac.in*

**Deterministic optical manipulation of fluorescent nanodiamonds (FNDs) in fluids has emerged as an experimental challenge in multimodal biological imaging. Designing and developing nano-optical trapping strategies to serve this purpose is an important task. In this letter, we show how chemically-prepared gold nanoparticles and silver nanowires can facilitate Opto-thermoelectric force to trap individual entities of FNDs using a long working distance lens, low power-density illumination (532 nm laser, 12 $\mu W/\mu m^2$). Our trapping configuration combines the thermoplasmonic fields generated by individual plasmonic nanoparticles and the opto-thermoelectric effect facilitated by the surfactant to realise a nano-optical trap down to a single FND 120 nm in diameter. We utilise the same trapping excitation source to capture the spectral signatures of single FNDs and track their position. By tracking the FND, we observe the differences in the dynamics of FND around different plasmonic structures. We envisage that our drop-casting platform can be extrapolated to perform targeted, low-power trapping, manipulation, and multimodal imaging of FNDs inside biological systems such as cells.**

Fluorescent nanodiamonds have garnered significant attention for their applications in sensing, biomedical imaging and quantum optics [1–4]. As a sensing material, FNDs have several unique advantages over other sensors as they are extremely stable, biocompatible, and can be engineered to respond to specific stimuli. They can also be used in harsh environments and detect a wide range of physical and chemical properties. In biomedical imaging, fluorescent nanodiamonds are used as labels to track the movement of cells and other biological structures. Their biocompatibility, small size, and bright fluorescence make them ideal imaging probes. They can also be functionalised with targeting molecules to selectively bind to specific cells or tissues, providing an even more specific imaging contrast. Overall, the unique properties of fluorescent nanodiamonds make them a promising material for sensing and biomedical imaging applications. But in order to expand their potential use for single spin imaging and optically detected magnetic imaging, their precise control in solutions is crucial[5–7]. While current studies on FNDs use separate laser beams for trapping and spectroscopy, this design can be limiting due to charge state perturbations and potential damage to sensitive environments such as living cells[8]. Thus, there is a need to develop new nano-optical and opto-thermophoretic trapping methods that facilitate trapping, spectroscopic probing and imaging using a single low-power laser[1,9–17].

Opto-thermoelectric trapping [18] has emerged as a promising technique for manipulating small particles, including nanoparticles and cells. A surfactant is added to the solution, which dissociates in anions and positively charged micelles. The particles to be trapped also get coated in a surfactant bilayer and become positively charged. A localised laser heating of a plasmonic structure (usually thin gold films) generates a thermal gradient leading to the separation of charges, as chlorine anions have higher mobility than the cationic micelles. This charge separation creates an electric field towards the heat source, creating an opto-thermoelectric force and trapping the positively charged particle. This technique has shown great potential for applications in biophysics, nanotechnology, and microfluidics, as it enables non-invasive and precise manipulation of particles in solution. In recent years, much research has been focused on exploring the potential of opto-thermoelectric trapping using various nanostructures, such as gold films, plasmonic nano-antennas, and metallic nanoparticles[18–20].

However, conventional opto-thermoelectric trapping using gold films has limitations, including low trapping efficiency, poor spatial control, and thermal damage to the trapped particles. The localisation is often improved by making the laser focus tighter, which leads to a restriction of having a short working distance from the objective. We previously showed low-power trapping of single gold nanoparticles using a plasmonic-nanostructure-based trapping platform[21]. This paper presents an approach for long working distance, low power opto-thermoelectric trapping of fluorescent nanodiamonds using plasmonic nanostructures instead of gold films or lithographically prepared structures.

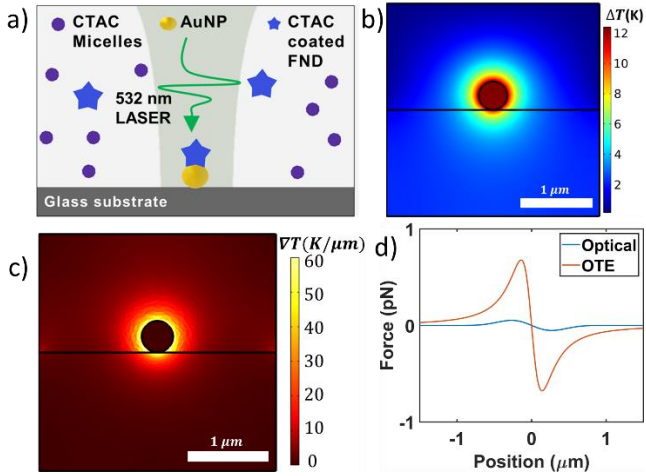

Fig. 1. a) Schematic of plasmon-assisted Opto-thermoelectric trapping of fluorescent nanodiamonds on Gold Nanoparticles. The surfactant (CTAC) facilitates an electric field which attracts the FND to AuNP. b) & c) Simulated temperature and temperature gradient distributions for a 400 nm AuNP placed on a glass substrate and surrounded by water. d) Comparison of opto-thermoelectric (OTE) and optical force in our geometry showing OTE force dominates the trapping.

We demonstrate that chemically synthesised plasmonic nanostructures can be used to generate a highly localised thermal gradient that can trap and manipulate fluorescent nanodiamonds with high efficiency and spatial resolution. To this effect, we used drop-casted plasmonic nanostructures, such as gold nanospheres (AuNPs) and silver nanowires (AgNWs). Figure 1a shows the schematic of our trap. The AuNPs could, in principle, be drop-casted on a desired substrate, and FNDs can be trapped there.

We investigated the capability of such anchor-particle-based trapping by calculating the OTE force on the FND. The temperature distribution around the nanoparticle was calculated using numerical calculations based on the finite element method using COMSOL Multiphysics (version 5.5) as a solver. The Wave optics module was coupled with the Heat transfer module to simulate the temperature increment and temperature gradient upon excitation with a light source. The simulation model consists of a 400 nm AuNP placed on a glass substrate and immersed in an aqueous solution. A focused Gaussian laser source of wavelength 532 nm, which is near the absorption maxima of the AuNP and FND, was used to excite the nanoparticle. This creates a temperature gradient around the AuNP due to the thermoplasmonic heating of the nanoparticle[22]. The size of the excitation spot ($3.34\ \mu m$) was taken from the experimental measurements. The resulting temperature gradient has been plotted in figure 1b. It can be observed that the temperature gradient is sharp in the vicinity of the nanoparticle and was found to be around 30 K/μm. From this thermal gradient, the thermoelectric field in our geometry was calculated as [20]

$$E_T = \frac{k_B T \nabla T}{e}\left(\frac{\sum X_i c_i S_{T,i}}{\sum X_i^2 c_i}\right)$$

Here i denotes the ionic species, $k_B$ is the Boltzmann constant, T is the surrounding temperature, e is the elementary charge, and $c_i$, and $S_{T,i}$ are the concentration and Soret diffusion coefficient of the ionic species i, respectively, and $X_i = \pm 1$ for positive and negative ions, respectively. See supplementary information for details on the calculation. The thermoelectric force, $F_{T,E}$, is then calculated as $F_{T,E} = \int \sigma E_T dA$, where $\sigma$ is the effective surface charge density of the FNDs measured indirectly using zeta-potential measurements. The thermoelectric force on the FND is calculated at the height of 10 nm above the AuNP. The optical force on the same particle is calculated using the Maxwell stress tensor method (refer supplementary information fig S2-4.) Figure 1c compares the OTE force with the optical force obtained for the same geometry. It is evident that the OTE force is dominant compared to the optical force, and therefore the resultant trapping potential is mainly facilitated by the OTE process.

Experiments. AuNPs and FNDs were purchased from Sigma Aldrich. The reported mean diameters of AuNP and FND are 150 nm ± 15 nm and 120 nm, respectively. AuNPs were redispersed in ethanol solution. The FNDs contain 3 ppm NV centres. FNDs were dispersed in a solution containing a specific concentration of cetyltrimethylammonium chloride (CTAC) molecules. AgNWs of an average diameter of 350 nm were synthesised using a polyol process[23] and immersed in an ethanol solution.

An assembly similar to the schematic of the AuNP driven opto-thermoelectric trap of FNDs in a microchamber shown in figure 1a was made. AuNPs dispersed in ethanol were drop-casted on a glass substrate, and the solvent was evaporated. These AuNPs remain firmly attached to the surface and do not re-diffuse back in the solution. A dilute solution of FNDs and surfactant (CTAC) is added to the drop-casted layer, and the chamber is sealed with an adhesive spacer and another glass coverslip. The chamber is placed on the stage of our custom-built optical trapping microscope, whose optical schematic is shown in figure 2. The L1 and L2 are lenses used to expand the beam to overfill the back aperture of the lens. The WL1 and WL2 are white light sources for top and dark-field illumination, respectively. The L3 and L4 lenses illuminate the sample plane with WL1, whereas a 1.4 NA darkfield condenser focuses white light from below. BS1 is a beam splitter combining white light and laser light in the input path of the upright microscope. A 50×, 0.5 NA microscope objective focuses the incoming beam on the anchored AuNP. The scattered light is collected by the top objective and is passed through a 532 nm notch filter to remove the Rayleigh scattered light. Then the emission is sent to a CCD camera, an EMCCD, or the spectrometer using mirrors M2 and M3. Lenses L5 and L6 focus the light to form the image on cameras and the spectrometer. This upright microscope has the following capabilities: darkfield imaging to visualise the drop-casted AuNPs (see supplementary figure, S6); 532 nm laser illumination to trap and excite the fluorescence in FND. The objective lens used for this purpose was a low numerical aperture (NA=0.5) air objective lens, ensuring that the power density delivered at the sample is low. The emitted fluorescence from a trapped FND was either captured by the spectrometer or directly imaged using an EM-CCD or a conventional CCD camera after rejecting the Rayleigh scattered light at 532 nm. This dual camera capability was harnessed to identify individual AuNP, trap, and perform single FND fluorescence tracking. The images were captured in the CCD camera at a frame rate of 35 frames per second for full frame and on the EMCCD at a frame rate of 60 fps. The position of the FND is tracked using the TrackMate[24] plugin available through Fiji[25].

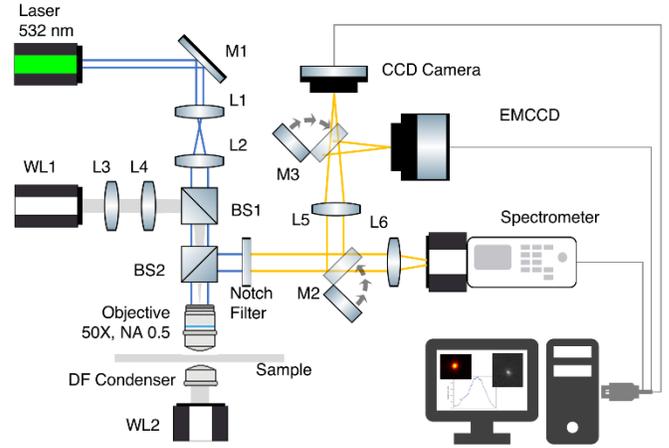

Fig. 2. Optical schematic of the trapping setup including lenses (L), mirrors (M), beam splitters (BS), white light sources (WL), 50x objective lens of 0.5 numerical aperture, a dark field (DF) condenser of numerical aperture 1.4, a charge-coupled device (CCD) camera, an electron multiplying CCD (EMCCD) camera, and a spectrometer. This scheme integrates dark-field microscopy, optical spectroscopy, tracking, and fluorescent imaging.

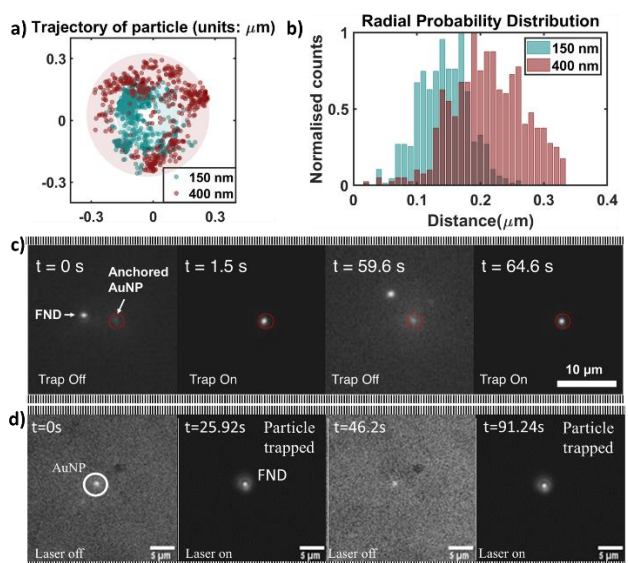
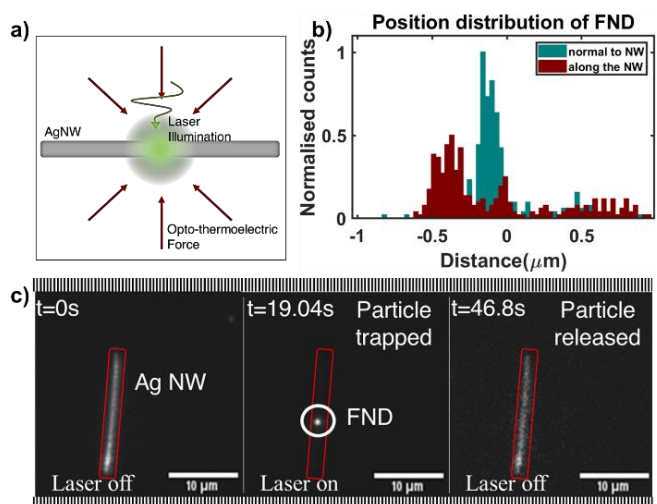

Fig. 3. a) Trajectories of trapped FND using 150 nm and 400 nm diameter AuNPs. The corresponding radial probability distribution shown in c) shows that the trapped FNDs radial distribution varies with the diameter of the anchor particle. Snapshots of single FND trapping are shown for c) 150 nm AuNP (See Visualization 1 and 2), d) 400 nm AuNP (See Visualization 3). The times series fluorescent images show the trap and release ability of a single FND.

Fig. 4. a) Schematic showing the trapping on Silver Nanowires (AgNW). b) Position distribution of FND is shown along the wire and in the direction perpendicular to the wire. c) Snapshots of single FND trapping on 400 nm diameter AgNW. The times series fluorescent images show the trap and release ability of a single FND (See visualization 4).

We next discuss the FND trapping in our AuNPs trap. We executed the trap using two different-sized AuNPs (150 nm and 400 nm in diameter). The sample trajectories for the trapped FND are plotted in figure 3a. In figure 3c, d, we show the fluorescence image of a reversible trap of a single FND achieved by controlling the illumination of AuNPs. Upon illumination of a single AuNP, we observed directed diffusion of FND towards the illuminated AuNP. When the laser is off, the FND diffuses back to the solution. The observed switching of the FND trap was relatively quick (about 10 seconds) and could be achieved repeatedly over multiple cycles. With the 150 nm AuNP based trap, the minimum laser power which could be used to trap the FND was 1.2 mW in the sample plane, corresponding to a power density of $34 \, \mu W/\mu m^2$. The absorption cross-section of a 400 nm particle is four times that of a 150 nm particle at 532 nm wavelength (see fig S1). Consequently, the minimum laser power with which trapping is achieved is 0.43 mW which implies a power density of $12.2 \, \mu W/\mu m^2$. This is much lower power than is required for trapping using conventional optical trapping (see supplemental figures S7,8). The plotted trajectory in figure 3a shows that the position of the trapped particle is not gaussian distributed as it is for optical trapping. The trap gets affected by the presence of AuNP in the centre. The particles diffuse around the AuNP, showing interesting dynamics as power and surfactant concentration are varied. The radial probability distribution of the trajectories plotted in figure 3b also shows the same. To plot the radial position distribution, the radial distance of the FND in the trap for all positions in the trajectory (shown in figure 3a) was calculated and then plotted as a histogram as a function of radial distance. Figure 3b shows the most probable radial position variation for the two different AuNPs, 130 nm and 210 nm, respectively, for 150 nm and 400 nm AuNP. Thus, we can manipulate the localisation of the trapped FND. The concentration dependence of the surfactant on the trajectory of the trapped particle is also studied and shown in the supplementary information, figure S5.

We also performed trapping experiments with 350 nm average diameter AgNWs used as the heat source. We have previously shown that these nanowires can be used as nanophotonic waveguides[26–28]. The schematic for the experiments is shown in figure 4a. The AgNWs were drop-casted on the glass substrate, and ethanol was evaporated. The aqueous solution of FND and CTAC was placed on top of it and sealed using an adhesive spacer and a coverslip. The trapping with AgNW is similar to the procedure described for the AuNP. Again, we used the laser polarized along the nanowire to excite the plasmons (and FND fluorescence) and create a thermal gradient. The snapshots of the trap are shown in figure 4c, where subsequent images show the trapping and release ability of the trap. The power required to trap the particle is, in this case, 3 mW ($85 \, \mu W/\mu m^2$). The position distribution of FND is shown parallel and perpendicular to the wire. It is observed that the particle has a broader position distribution parallel to the wire than in the perpendicular direction. This can be understood as the temperature distribution is much steeper in the perpendicular direction to the nanowire than in the parallel direction, as we have previously shown[28].

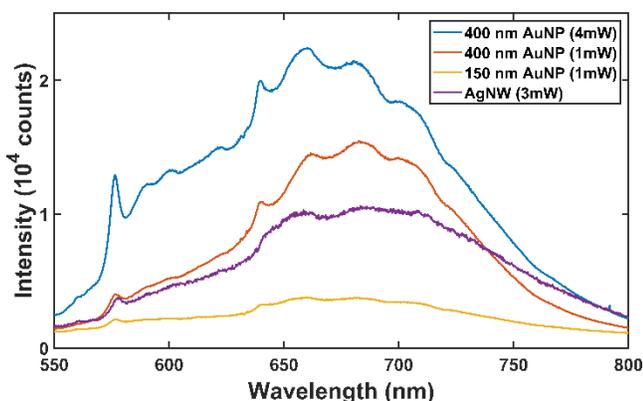

Fig. 5. Fluorescent spectra collected for 1 second from a single FND trapped in our trap on different plasmonic structures. The zero-phonon line at 575 nm and the NV- line at 638 nm are marked, demonstrating the trap's capability to investigate the trapped particle's spectral features.

The trapped FND was spectroscopically probed for all three implemented traps, as shown in Figure 5. In the recorded spectra, we observed the zero-phonon line at 575 nm and the NV- line at 638 nm, which indicates the charge state. These assignments fit well with the literature[2,29]. This further suggests that the spectral features of FND are preserved in our plasmonic nanostructure-based OTE trap. Such heterodimer systems can potentially read out local physical perturbation and chemical reactions in and around the illuminated plasmonic nanostructures.

**Conclusion.** Optical trapping of nanoparticles such as FNDs at minimum laser power has implications in multimodal imaging of biological systems such as cells and tissues. Facilitating sufficient trap stiffness to manipulate FNDs with spectroscopic imaging capability is an experimental challenge. Our work shows a pathway to achieve optothermal trapping and spectroscopy with a single excitation source (532 nm) at low laser power density (around 12.2 $\mu W/\mu m^2$) and long working distance by using a low numerical aperture lens. Specifically, we have shown that trapping can be achieved with chemically prepared plasmonic structures instead of lithographically prepared gold films or gold nano-antennae. The realised gold nanoparticle, and silver nanowire-based traps combine optical trapping with thermophoretic and thermoelectric effects and serve as a platform for nano-manipulation in various fluid environments. We also show variations in the dynamics of particles in trapping experiments with different plasmonic structures. By adding complementary fields such as radio frequency and weak magnetic perturbations, we envisage that our low-power trapping method can be extrapolated to optically detect magnetic resonance and single spin imaging inside a biological cell.

**Supplementary material.** Visualization showing the dark-field visualisation and trapping ability of the trap and the ability of the plasmonic structures driven OTE trap to trap and release a single nanodiamond.. Supplemental document discussing: Description of the videos, Mechanism of trapping, calculated absorption of gold nanoparticles, the zeta potential of nanodiamonds, details of opto-thermoelectric force calculation, the dark-field image of gold nanoparticles, comparison of optical and opto-thermoelectric trapping, and variation of trapping as a function of concentration for the AuNPs. (Pdf)

**Funding.** G.V.P. acknowledges financial support from the Air Force Research Laboratory grant (FA2386-18-1–4118 R\&D 18IOA118) and the Swarnajayanti fellowship grant (DST/SJF/PSA-02/2017–18). A.S. acknowledges CSIR, and PMRF for the Junior research fellowship and PMRF. K.S. acknowledges financial support from DST Inspire Faculty Fellowship DST/INSPIRE/04/2016/002284 and DST Quest Grant DST/ICPS/QuST/Theme-2/2019/Q-58.

**Acknowledgements.** We thank Rahul Chand, Eksha Rani Chaudhary, and Diptabrata Paul for their valuable support and feedback.

**Disclosures.** The authors declare no conflicts of interest.

**Data availability.** Data underlying the results presented in this paper are not publicly available at this time but may be obtained from the authors upon reasonable request.

# OPTO-THERMOELECTRIC TRAPPING OF FLUORESCENT NANODIAMONDS ON PLASMONIC NANOSTRUCTURES: SUPPLEMENTAL DOCUMENT

**Media** Description of the videos.

Visualization 1. Individual FND diffuses towards the anchored 150 nm Au np, which is at the center of the screen. The FND gets trapped and the trap stays active for about 30 seconds. The laser is then switched off to allow the FND to diffuse away. The dark field illumination is then turned which shows multiple FNDs floating around and four drop-casted Au nps dispersed on the sample plane. The position of the Au np is checked to be at the laser center and then the trap is turned on after turning the dark-field illumination off. This is done to ensure that only illumination from FNDs is captured, which helps in accurate tracking and spectral information processing. The FND is readily trapped as shown. The video is recorded at 20 mM concentration of CTAC and 4 mW laser power.

Visualization 2. A single FND is visible at the beginning of the video, which is diffusing towards the anchor 150 nm Au np. Its fluorescence intensity increases as it approaches trap center. The total intensity is scaled for better visibility. The FND gets trapped and stays in trap for about 5 seconds. The laser is then switched on and off to allow the FND to diffuse away. It is again recaptured and stays in the trap for as long as the trap is active. The videos are recorded at 20 mM concentration of CTAC and 4 mW laser power.

Visualization 3. FND is repeatedly trapped and released using 400 nm Au np as the anchor particle. The video is recorded at 14 mM CTAC concentration and 1.2 mW laser power.

Visualization 4. FND is trapped and released using (350 nm edge-to-edge thickness) AgNW. The video is recorded at 14 mM CTAC concentration and 3 mW laser power.

**Calculated absorption spectra of Au nps**

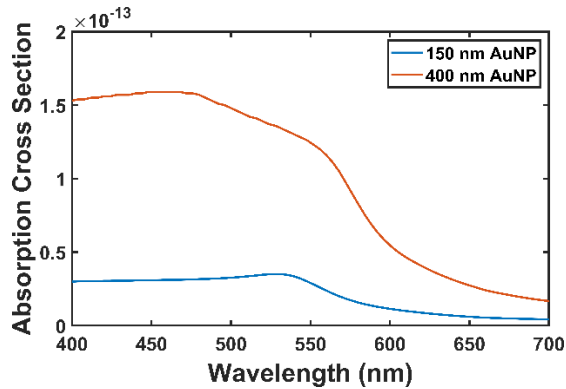

Fig. S1. Calculated absorption cross section of the drop-casted Au np (150 and 400 nm diameter).

The absorption cross section was calculated using this effective refractive index (n = 1.365) to represent Au np on a glass substrate surrounded by water. This effective refractive index was estimated by matching the experimentally measured scattering cross section of the Au nps with calculated cross section of a 150nm Au np [1,2]. The absorption cross section for 400 nm Au np is nearly 4 times the cross section of 150nm Au np. Thus, it can generate more heat for lower laser power.

**Mechanism of reversible trapping of FND using single nanoparticle driven opto-thermoelectric trap.**

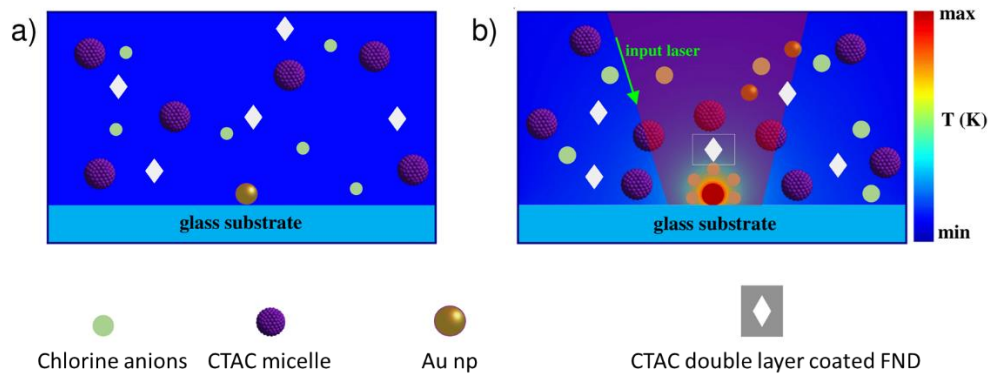

Fig. S2. Mechanism of reversible trapping of FND using single nanoparticle driven opto-thermoelectric trap. a) Gold particles are drop-casted and aqueous solution of fluorescent nanodiamonds with surfactant is added. b) Laser on Au np creates a thermal gradient driving the trapping.

The figure S2 shows the mechanism of reversible trapping. Au nps are drop-casted on glass surface and dried in a desiccator. A chamber is prepared in which the solution containing FNDs, and CTAC molecules of desired concentration is put. The chamber is then sealed using another glass slide. The CTAC molecule dissociates into Chlorine ion and the corresponding positively charged ion. The long chained positive ion coats the FNDs and the Au nps with a bilayer making them positively charged. The solution also contains micelles of the CTAC.

When the Au np is illuminated with laser, a temperature gradient is set up in the surrounding fluid. Due to difference is the Soret thermos-diffusion coefficient of the positive CTAC micelles and the chlorine ions, a separation of charge occurs. The CTAC micelles, larger in size, have a higher Soret coefficient and move farther away from the heat source while the chlorine ions do not move that far away as shown in Figure S2 (b). This creates an electric field which pulls any positively charged particle towards the heat source. The trap can be turned off by turning the laser off as the charges get redistributed when the temperature gradient is removed.

**Zeta potential.** Measured Zeta potential of FNDs before treating them with CTAC.

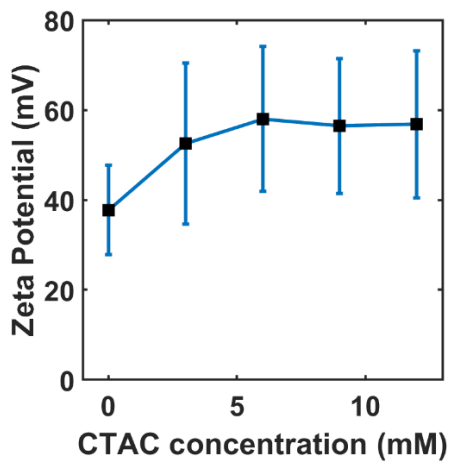

Fig. S3. Zeta potential of FNDs plotted as a function of concentration of CTAC.

The zeta potential of the particles was measured for various concentrations using dynamic light scattering experiment. The surface charge from this zeta potential is calculated by using Gouy-Chapman relation [3].

**Force calculation**

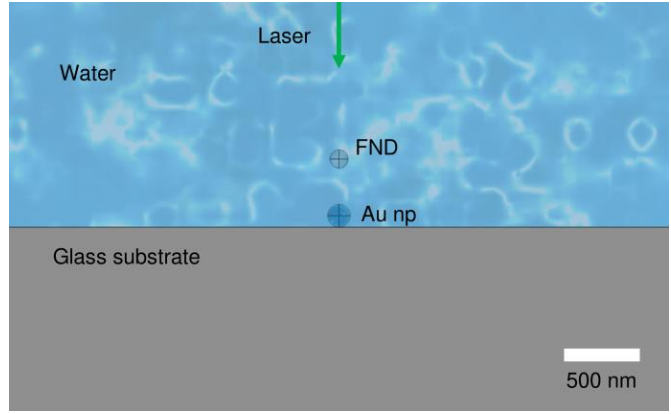

Fig. S4. Snapshot of geometry used for numerical calculation of the temperature gradient around the particle.

As mentioned in the main text, the thermoelectric force on the FND was calculated using the following equations:

$$E_T = \frac{k_B T \nabla T}{e} \left( \frac{\sum X_i c_i S_{T_i}}{\sum X_i^2 c_i} \right)$$

Here i denotes the ionic species, $k_B$ is the Boltzmann constant, T is the surrounding temperature, e is the elementary charge, and $c_i$, and $S_{T,i}$ are the concentration and Soret diffusion coefficient of the ionic species i respectively, and $X_i = \pm 1$ for positive and negative ions respectively.
The thermoelectric force, $F_{T,E}$, is then calculated as

$$\boldsymbol{F_{T,E} = \int \sigma E_T dA}$$

where $\sigma$ is the effective surface charge density of the FND measured indirectly using zeta-potential measurements. The $S_{T,\ CTAC}$ and $S_{T,\ Cl}$ are taken from references [4,5] respectively.

The force profile shown in Figure 4b is calculated by scanning the FND along x direction at a height of 10 nm above the glass surface. The optical force was calculated on the same assembly by integrating the Maxwell's stress tensor [6] on the surface of the FND. The refractive indices of materials were taken from references [7,8]. Spherical particles were assumed and no thermal effects were included while calculating the optical force.

**Concentration dependence of Trapping**

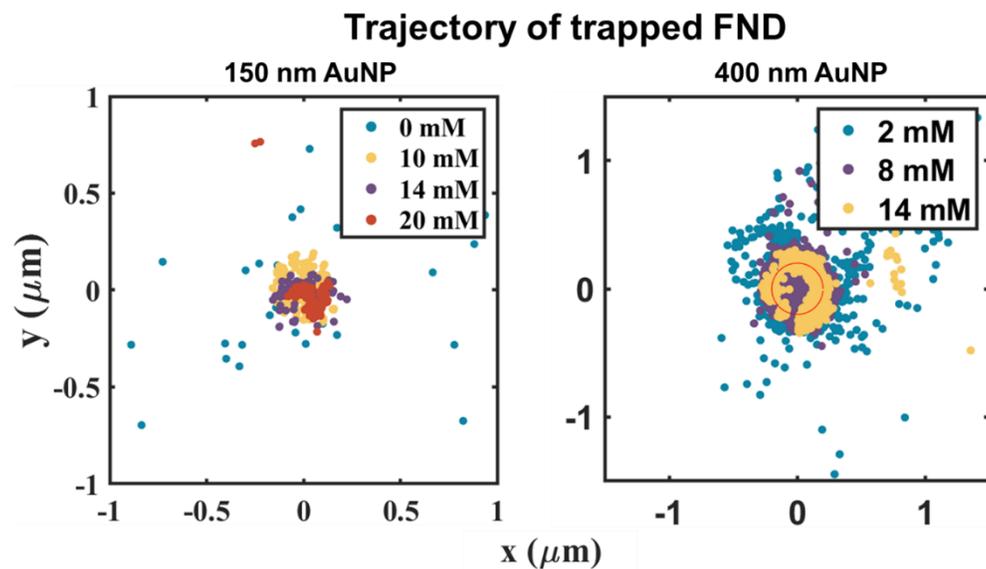

Fig. S5. The trajectories of particles are shown for two spherical anchor particles (150 nm and 400 nm Au np). The particles are trapped more stiffly as concentration is increased.

The figure S5 shows variation of trajectories of particles as the concentration is varied. The increased concentration of CTAC creates stronger electric field that leads to stiffer trapping. The variation is showed for 150 nm and 400 nm. The trapping power in both cases is 1.1 mW and 1.2 mW respectively.

**Dark field image of the gold particles**

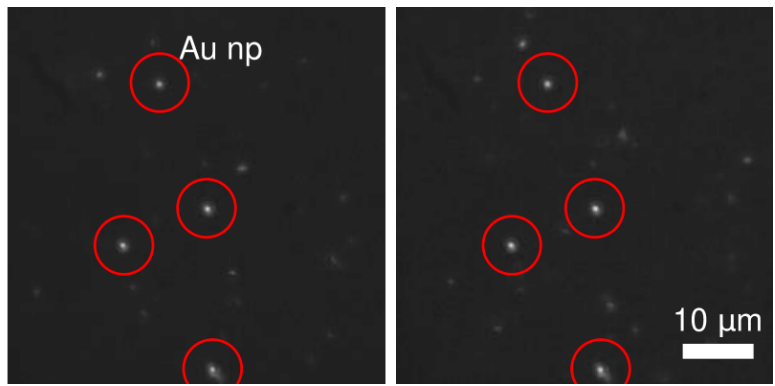

Fig. S6. Dark field image (captured in EMCCD) showing 150 nm Au nps with fainter FNDs floating around. The two panels of the image show drift in position of FNDs while the anchor Au nps stay in place.

**Comparison of Optical trapping and Opto-thermoelectric trapping**

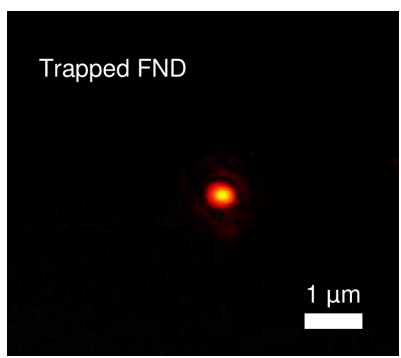

Fig. S7. Fluorescence image of optically trapped multiple FNDs.

The nanodiamonds were optically trapped in same setup using an objective lens of 0.5 numerical aperture. The trapping is done at a relatively higher concentration. Thus, multiple particles come in the beam spot and are trapped. The power required to stably trap the FNDs this way in the setup exceeds 20 mW.

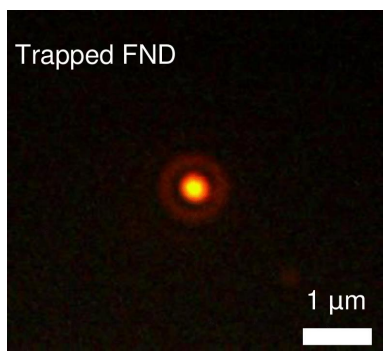

Fig. S8. Fluorescence image of single optically trapped FND using a water immersion objective lens (NA 1.2).

To maximize the trapping efficiency for optical trapping we used an objective lens with high numerical aperture to increase the gradient force. The input power required to trap the particle is more than 15 mW with this setup.